\documentclass[12pt]{article}
\usepackage{epsfig}
\usepackage{amssymb}

\newcommand{\kB}{k_B}
\newcommand{\summa}[2]{ \sum_{ #1 \: #2 }}
\newcommand{\surf}{^{\textrm{\scriptsize (s)}}}
\newcommand{\s}{q}

\author{U.~Tartaglino$^{a,b,\dag}$, D.~Passerone$^{d,a,b}$,
        E.~Tosatti$^{a,b,c}$ and F.~Di Tolla$^{a,b}$}
\title{Bent surface free energy differences from simulation}

\date{September 29, 2000}

\begin{document}

\maketitle

\noindent {\it
$^a$ SISSA, Trieste, Italy \\
$^b$ Unit\'a INFM/SISSA \\
$^c$ ICTP, Trieste, Italy \\
$^d$ Max-Planck-Institut f.~Festkoerperforschung, Stuttgart, Germany \\
$^\dag$ Corresponding author. Address: SISSA, Via Beirut 2, I-34014
  Trieste, Italy. FAX: +39-040-3787528. E-mail: tartagli@sissa.it
}
\vspace{1em}

\noindent
ECOSS 19 -- Abstract number: 00456

{
\abstract
We present a calculation of the change of free energy 
of a solid surface upon bending of the solid. It is based 
on extracting the surface stress through a molecular dynamics 
simulation of a bent slab by using a generalized stress
theorem formula, and subsequent integration of the stress with respect 
to strain as a function of bending curvature. The method is exemplified 
by obtaining and comparing free energy changes with curvature
of various reconstructed Au(001) surfaces.}
\vspace{1em}


\section{Introduction}

The surface free energy of a solid, defined as the excess free 
energy of the semi-infinite solid relative to an infinite solid
with equal number of particles\cite{Landau},
is a very important but also a very elusive quantity. 
It is important because its minimum
determines the surface equilibrium state and its properties. However,
while in the liquid it coincides with the ordinary surface tension, 
in the solid it becomes elusive because there it can neither be easily 
measured nor calculated. The quantity which is easy to access is instead
the surface stress (generally a rank 2 tensor) defined as the change of the
surface free energy with respect to a unitary increase of surface area
upon stretching, which combines surface specific free energy $\gamma$
and its first derivative:
\[
 \sigma\surf = \gamma + A \frac{d\gamma}{dA}
\]
While $\gamma$ is of course positive for a substance below its critical
point, the magnitude of the second term (arising from
the rigidity of the solid) is generally comparable, but can have
either sign. Therefore, while surface stress is becoming increasingly
available through measurement\cite{Ibach}, simulation\cite{RibarskyLandman},
and microscopic calculation\cite{Martin},
the surface free energy remains generally unknown.
At $T=0$, where surface free energy and surface energy coincide,
there are of course a large number of calculations, as quoted e.g.\
in Ref. \cite{Ibach}. However, finite temperature free energy
calculations, including the entropy term, seem totally missing.
This is particularly
frustrating in view of the desire to characterize and possibly
predict surface phase transitions as a
function of parameters, such as temperature, coverage, external
stresses, crystal bending, etc.  In systems where interatomic forces
are well understood, an alternative methods to predict
such phase transition is simulation, particularly  Molecular Dynamics 
(MD) simulation. However, besides being somewhat less fundamental,
the simulation approach usually suffers from practical problems, such
as size and time limitations, which greatly restrict the variety of
transitions that can be directly described.

In the present paper we present a route to calculate directly
the surface free energy {\em change}
with respect to one specific external parameter, the
curvature of the underlying solid. The method is again 
based on Molecular Dynamics (MD) simulations, and uses the framework
for variable curvature MD developed by Passerone et al.\cite{Passerone}.
The surface phases to be compared can be either simulated separately, or
realized on the two opposite faces of the same simulation
slab. Since free energies are calculated separately, the actual phase
transformation is not required to take place spontaneously
in the simulation -- in fact, it must be avoided.
Curvature of a crystal plate is a standard tool for observing
surface stress difference of the two opposite faces of the
plate\cite{KochAbermann,Ibach}. Conversely, inducing curvature of the 
plate by an external bending force will cause the two opposite
faces to depart from their original state, and their free energies
will generally evolve in opposite directions.

We have chosen metallic surfaces as our test case, for a variety
of reasons. The surfaces of metals are readily accessible to
a variety of surface techniques. Especially noble and some
transition metals show a variety of surface reconstructions
where the surface stress plays an important role\cite{Ibach}.  
If a metal slab is bent the surfaces are strained and
work is exerted onto them. It is plausible that bending could
eventually drive phase transitions, including changes or removal of 
the reconstruction. Preliminary results on reconstructed
Au(111) indicate that the surface reconstruction, 
or at least some features of it, can in fact be removed by 
curvature\cite{Zeppenfeld}

\section{Method}

Although MD is a very useful technique to study the time
evolution of energies, temperature, pressure, stress, \ldots\  of
a system with well defined forces, other
thermodynamic quantities, like the free energy and entropy,
are not obtained. The reason is that such quantities are not an average
of some mechanical entity over the ensemble of the states: 
they contain information on the whole ensemble of
states which cannot be directly extracted from the time evolution 
of a sample\cite{Frenkel}. Nonetheless in some specific cases 
the free energy variation along a reversible, isothermal path 
can be determined from mechanical quantities by integrating the 
work done onto the system.

In the particular case of the slab, where curvature is forced 
externally through bending, the work is given by the
integral of the stress tensor with respect to the strain and the volume
of the slab. This integral contains contributions both from
the surfaces and from the bulk of the slab. For our purposes,
the former must be separated from the latter\cite{Nota1}.
The first step is to extract the total stress field 
$\sigma_{\alpha\beta}$ in the simulation of a curved slab.
The only relevant component is $\sigma_{yy}$, where
the $y$-axis is parallel to the surface and is oriented in the direction
that is stretched during bending (the other axes are chosen with $z$
normal to the slab). The average force exchanged across the $xy$-plane
is\cite{RibarskyLandman}
\begin{equation}
\label{Tyy}
 T_{yy} \equiv \sigma_{yy}\cdot L_x L_z =
 - \frac{1}{L_y}\left\langle \sum_i \frac{p_{iy}^2}{m_i} \right\rangle
 - \frac{1}{2 L_y}\left\langle
    \summa{i,j}{(i\neq j)}
    \frac{\partial U}{\partial r_{ij}} \frac{(x_i-x_j)^2}{r_{ij}}
   \right\rangle~,
\end{equation}
where the indexes $i$ and $j$ runs over the particles, $U$ is the potential
energy as a function of the interatomic distances
$r_{ij}\equiv |\vec{r}_i-\vec{r}_j|$ and $L_x$, $L_y$, $L_z$
indicate the lengths of the slab along the three axes (the latter being
the slab thickness). Note that this formula does not require pairwise 
potentials as it might seem. It applies for arbitrary many body potentials 
(including EAM, glue models, etc) such that total energy depends 
only on pair distances.
In the following for convenience we shall write $F_{ij}$ in place of
$\partial U/\partial r_{ij}$. Note that this is generally not the force 
acting between particles $i$ and $j$, but it becomes that in the specific 
case of pairwise additive interactions.
The first term in the RHS is the kinetic energy contribution to stress,
which in the classical case is just
$(N \kB T / V) \cdot L_x L_y$, being $N$ the number of atoms
and $V\equiv L_x L_y L_z$ the volume of the cell.
The second term arises from interparticle interactions:
it is the derivative of the potential energy with respect to
a uniform stretching along the $y$-axis, averaged over the system
configurations. 

We cut a slab in slices along the $z$ direction, each slice
corresponding to one layer (the depth coordinate $z$
is orthogonal to the bending axis $x$, and to the bending direction $y$).
By restricting the sum to particles located in each slice
we calculate the stress along the bending direction, resolved layer by layer. 
For a bent slab, equation (\ref{Tyy}) must be generalized 
to compute the average stress per layer in the
direction which locally follows the profile of the surface.
This is done by computing the potential energy derivative
for a uniform strain of the slab along y.
In order to better deal with the symmetry of the bent slab,
we introduce a system of scaled, curved coordinates $(\s_1,\s_2,\s_3)$;
indicating with $k$ the slab curvature in the neutral cylinder (i.e.\  in
the middle of the slab):
\begin{equation} \label{coordinates}
 \left\{
  \begin{array}{l}
   x = L_x \s_1 \\
   y = (1/k+L_z \s_3) \sin (k L_y \s_2) \\
   z = (1/k+L_z \s_3)\cos(k L_y \s_2) - 1/k
  \end{array}
 \right.
\end{equation}
The variable $\s_2$ is proportional to the polar angle $\theta$ through
$\theta=kL_y\s_2$; thus $\s_2$ moves along the bending direction.
The ``radial'' coordinate $\s_3$ is measured along the
direction normal to the surface and is zero in the middle layer.
Each of these coordinates ranges between $-1/2$ and $1/2$.
In the limit of zero curvature they are proportional to $x$, $y$
and $z$ respectively and correspond to the scaled coordinates
introduced by Andersen\cite{Andersen} and by Parrinello and
Rahman\cite{PR}.
We note that this system of coordinates is slightly
different from the one introduce in ref.\ \cite{Passerone}:
the present choice is more convenient in order to compute
the stress field.
Omitting the details of the calculation, the final formula
for the derivative of the potential energy keeping all the scaled
coordinates fixed is
\begin{equation}
 \label{kderiva}
 \frac{\partial U}{\partial L_Y} =
   \frac{1}{2} \summa{i,j}{(i\neq j)}
    \left[
    \frac{F_{ij}}{r_{ij}} (1+\s_{3i} k L_z)(1+\s_{3j} k L_z)
     \frac{\theta_{ij} \sin (\theta_{ij}) }{k^2 L_y}
    \right]
\end{equation}
Since periodic boundary conditions are used, both the distances $r_{ij}$
and the differences $\theta_{ij}\equiv \theta_i-\theta_j =
k L_y (\s_{2i}-\s_{2j})$  must be computed with the
\emph{minimum image convention}.
The contribution of particle $i$ to the force
$\int\!\!\int\!dxdz\,\sigma_{22}$ is
\[
  - \frac{\kB T}{L_y}
  - \frac{1}{2}\summa{j}{(j\neq i)}
    \left[
    \frac{F_{ij}}{r_{ij}} (1+\s_{3i} k L_z)(1+\s_{3j} k L_z)
     \frac{\theta_{ij} \sin (\theta_{ij}) }{k^2 L_y}
    \right]
\]
The average force any layer exerts in the $\s_2$ direction
is obtained as:
\begin{eqnarray}
\label{T22}
  T_{22} & = &
  - \left\langle N_{\mbox{\scriptsize layer}}\right\rangle \frac{\kB T}{L_y}
  - \Bigg\langle \sum_{i\in\mbox{\scriptsize layer}} \Bigg\{
    \frac{1}{2}\summa{j\in\mbox{\scriptsize slab}}{(j\neq i)}
     \bigg[
    \nonumber \\
    & &
    \frac{F_{ij}}{r_{ij}}\, (1+\s_{3i} k L_z)\,(1+\s_{3j} k L_z)\,
     \frac{\theta_{ij} \sin (\theta_{ij}) }{k^2 L_y}\,
     \bigg]
     \Bigg\} \Bigg\rangle 
\end{eqnarray}
In the zero curvature limit , $k\to 0$, this formula reduces to 
(\ref{Tyy}).

\begin{figure}
\epsfig{file=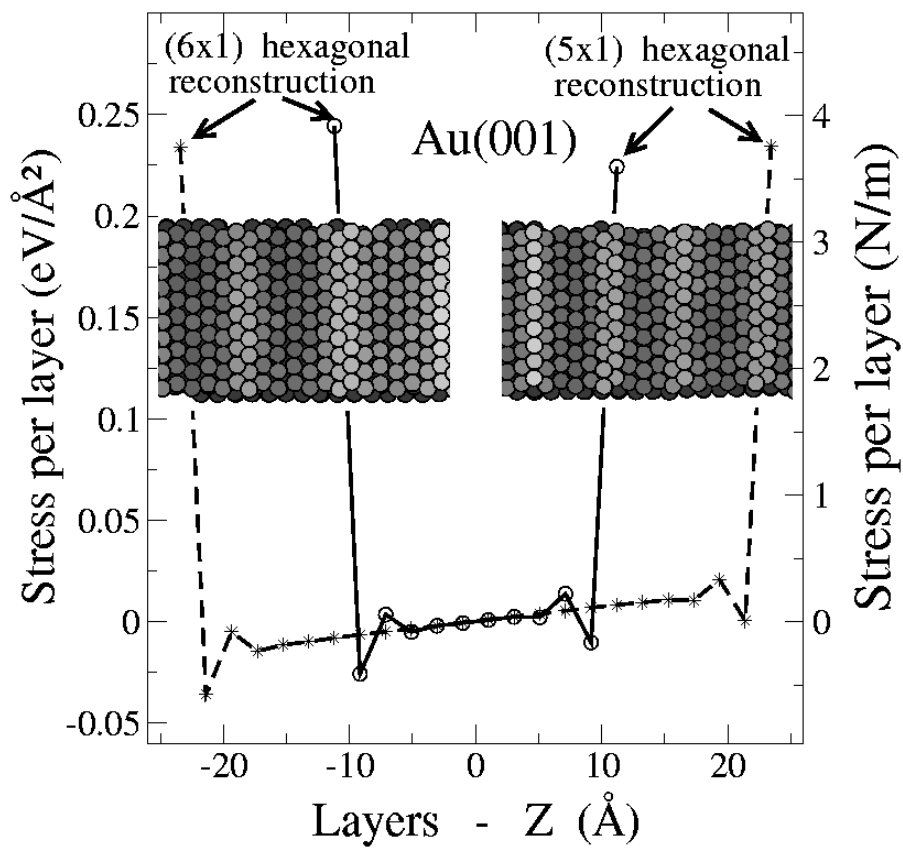, width=0.9\textwidth}
\caption{\label{stressperlayer} The linear stress (force/length) associated
to every layer of two samples with different thicknesses.
In the insets the reconstructions of the two surfaces are shown.
Surface atoms with the largest $|z|$ are shown as lighter, and identify
the reconstruction solitons. Note the bulk linear $z$ dependence,
and the large surface contribution.}
\end{figure}

To exemplify, we apply this scheme to the Au(001) surface,
modeled by means of the ``glue'' many body 
potential\cite{colla}. 
It is well known that this surface reconstructs 
with a denser triangular top layer\cite{review,expAu001},
increasing its lateral density 
by about 24\% relative to an unreconstructed layer.
One could expect that the tensile stress typical of the unreconstructed
surface should have decreased, maybe even disappeared, or reversed.  
Moreover, different reconstruction periodicities with different
lateral densities and different surface stresses might come
in competition with one another for the lowest free energy, as
curvature is cranked up. 
Figure \ref{stressperlayer} shows the stress  $T_{22}/L_x$
in each slice of a (001) gold slab. The two slab surfaces are given two
slightly different reconstructions, namely $1\times 5$ (close to the
experimental one) and $1\times 6$ (with slightly lower surface density).
Two slab samples are considered: one made by 12 layers and a thicker one
made by 24 layers. Both samples have sizes $L_x=28.8$\AA\ and
$L_y=172.9$\AA\@. The bending direction is [110].
The stress is computed through (\ref{T22}) taking the averages during
a 36 $p$sec evolution time at $T=100$ Kelvin.
The bending curvature in the middle plane $z=0$ where strain vanishes,
is $k=4.49\cdot 10^{-4}\textrm{\AA}^{-1}$.
The strain of such a curved slab
is linear with $z$ by construction and is determined by the curvature $k$
through the simple relation $\varepsilon_{22}=kz$.
The z-resolved stress distribution is instructive.
The stress profile deep in the slab is, as expected, again linear with $z$, 
and proportional to the strain. Close to the surfaces, however, the stress 
differs from its bulk-like extrapolation, and oscillates.
Both surface top layers show a positive (tensile) stress,
that is the surfaces further tend to reduce their area. This is rather 
the rule on metal surfaces\cite{Ibach}, but it is interesting to
note that even the more close-packed $1\times 5$ reconstruction has not quite
eliminated the tensile stress. 
The oscillations, in turn, reflect the composite layer-dependent
response to the main tensile 
force exerted by the top layer.
In order to calculate the {\em total} surface stress, we must subtract
from the actual stress distribution the corresponding extrapolated
bulk linear stress, and integrate.
The final result must be independent of the slab thickness.
Comparison of the slab-resolved stress of the thin slab with
that of the thicker one, as in
Figure \ref{stressperlayer}, shows that this is indeed
the case, and that the procedure works even in the thinner one. 
In the following we will write $T_{22}\surf(l)$ for the
surface contribution to the force associated to layer $l$.
The surface stress is
\[
 \sigma_{22}\surf = \frac{1}{L_x} \sum_l T_{22}\surf (l)~,
\]
where the sum extends over the layer close to the interface
with not negligibly small $T_{22}\surf$. The value obtained for
Au(100) is of $4.0$ N/m in excellent agreement with the experimental
value of $4.4$ N/m \cite{expstress}.

\section{Curvature-Dependent Free Energy}

Now we are ready for the next step in the free energy difference 
calculation. As the curvature is varied in an isothermal environment 
the reversible work done against the surface contribution should equal
the surface free energy variation. Since both strain and stress
depend on the layer depth ($\varepsilon_{22}(l)=kz_l$), the sum
must be done layer by layer
\[
 dF\surf =
    L_y \sum_l T_{22}\surf(l) \, d\varepsilon_{22}(l)
\]
Replacing the strain $\varepsilon_{22}(l)$ with $z_l k$ and integrating over
the curvature $k$:
\[
 \Delta F\surf = L_y \int dk \, \sum_l T_{22}\surf(l)\, z_l
\]
In conclusion the \emph{surface free energy per unit area} is linked to
its value $f_0\surf$ for the flat slab through the relation
\begin{equation}
 \label{fsurf}
 f\surf = f_0\surf + \frac{1}{L_x}\int_0^k dk \, \sum_l T_{22}\surf(l)\, z_l
\end{equation}

\begin{figure}
 \epsfig{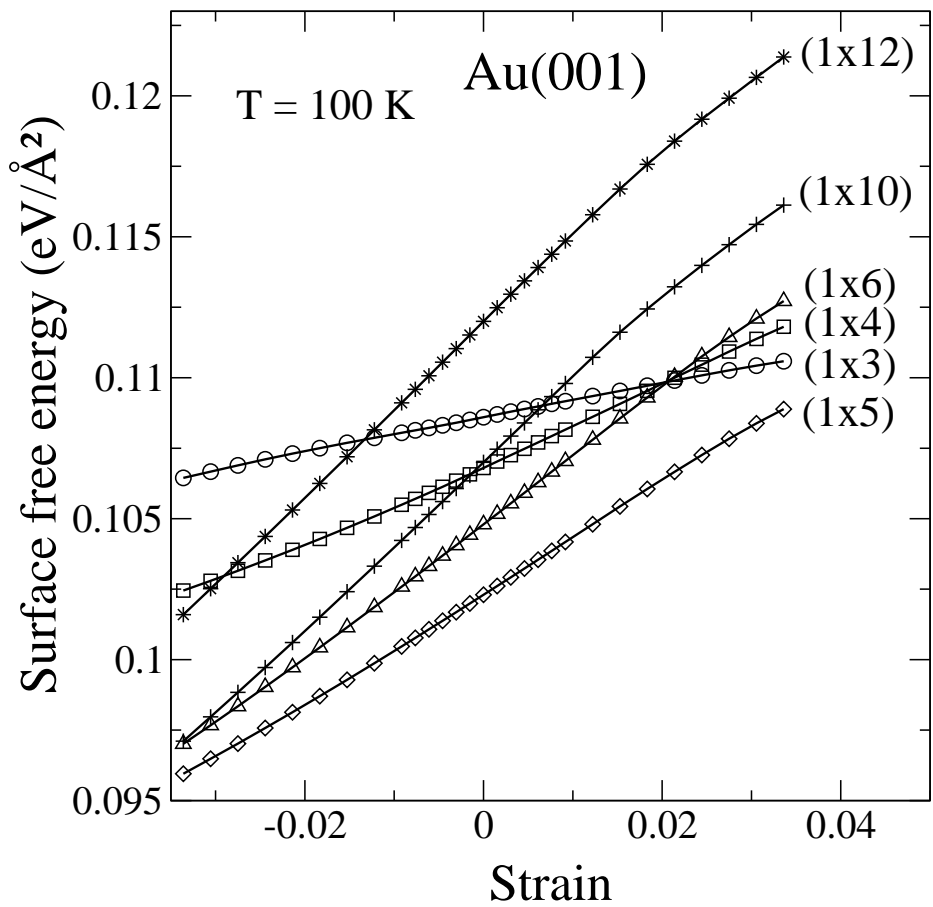}
\caption{ \label{ooifig} The free energy per unit area for
 some reconstructed Au(001) surfaces. In abscissa the strain at
 the surface due to bending is reported. }
\end{figure}

We continue our exemplification of the method
to the (001) surface of gold, where we wish to calculate curvature-dependent
free energies.
We prepared different (001) slab samples, 
where the surfaces differ by a different reconstruction periodicity,
in turn related to a different density of added atom rows perpendicular
to the bending direction, that is parallel to the $x$-axis.
In particular, in a $(1\times n)$ 
reconstruction one extra row is added every $n$ existing rows. 
After the usual equilibration steps the surfaces relax and 
equally spaced solitons appear, corresponding to
misfit dislocations between first and second layer.
The slabs are gradually bent and for
each curvature the overall layer-resolved stress is computed 
through (\ref{T22}), the averages taken during a 36 $p$sec
evolution time in a canonical simulation at $T=100K$. 
From the forces $T_{22}$ the surface
contributions are extracted, as explained above.
We plot the results in Fig. \ref{ooifig}, where we have for
convenience assumed arbitrarily the unknown zero-curvature free energy
to be equal to the zero-temperature energy, calculated previously
through MD by F. Ercolessi et al.\cite{ETP}.
At our working temperature of 100 K, this is likely
a not unreasonable approximation.

All the surface free energies decrease with curvature on the concave
side, where the compressive subsurface strain acts to reduce
the need for a tensile surface stress. The decrease is more pronounced, as
it should be, for the less dense higher reconstruction, whose
tensile stress is higher. Although there are crossings, no
other reconstruction crosses the $(1\times 5)$ surface free energy,
which is therefore
predicted to be stable against curvatures leading to surface
strains up to the percent range.

Preliminary as these results are, they seem encouraging, although
we do not yet have an alternative route to check them. It will
be interesting in the future to carry out tests aimed among other things
at separating the internal energy from the entropy part. 

In conclusion, we have presented a calculation of surface free
energy variation with curvature, realized by direct integration
of surface stress, obtained through  realistic molecular dynamics
simulation. A specific application to (001) reconstructed gold 
surfaces shows a good feasibility of the method, and foreshadows
future applications to study surface phase transitions under
curvature.

\begin{center}
 \textbf{\Large Acknowledgments}
\end{center}

We acknowledge support from MURST COFIN99, and from INFM/F.
Research of D. P. in Stuttgart is supported by the Alexander Von Humboldt 
foundation.



\begin{thebibliography}{99}
\bibitem{Nota1} We observe that
the curvature involves a non-uniform anisotropic strain, whence
the system as a whole is out of equilibrium: thermodynamic
arguments are only meaningful for the interface.
\end{thebibliography}
\end{document}